# Buckling and competition of energy and entropy lead conformation of single-walled carbon nanocones

Shengli Zhang[a]
*Department of Applied Physics, Xi'an Jiaotong University, Xi'an 710049, China and China Center of Advanced Science and Technology (World Laboratory), P.O. Box 8730, Beijing 100080, China*

Zhenwei Yao, Shumin Zhao, and Erhu Zhang
*Department of Applied Physics, Xi'an Jiaotong University, Xi'an 710049, China*

Using a continuum model, expressions for the elastic energy, defect energy, structure entropy, and mixing entropy of carbon nanocones are proposed analytically. The optimal conformation of carbon nanocones is studied by imposing minimization of free energy and analyzing the effects that the buckling of a nanocone's walls have during formation. The model explains the experimentally observed preference of 19.2° for the cone angle of carbon nanocone. Furthermore, it predicts the optimal conformation of carbon nanocones to result in a cone angle of 19.2°, radius of 0.35 nm, and critical length of 24 nm, all of which agree very well with experimental observations.

There has been dramatic progress in the synthesis and manipulation of carbon nanomaterial in recent years. Besides carbon nanotubes[1,2] and fullerenes,[3] scientists have observed other freestanding carbon nanostructures such as onionlike structures,[4] regular coils,[5] ring ropes,[6] and nanocones.[7–9] Because of their unique geometrical structure, carbon nanocones have many potential applications.[10–12] For example, Ganser *et al.* found that HIV-1 has the structure of a nanocone.[10]

Theoretically, the energies, geometries, and the electronic properties of tips of carbon nanocone have been studied.[12–17] According to Euler's theorem, the cone angle $\alpha$ is given by $\sin(\alpha/2) = 1 - P/6$, where $P$ stands for the number of carbon pentagons on the smaller cap and takes the values of 1, 2, 3, 4, 5, or 6. Experimentally it has been found that only the cone angle corresponding to $P=5$ occurs, but why this is so has not yet been resolved. Furthermore, the cone angle of most synthesized single-walled carbon nanocones is 19.2° (Ref. 9), while upper limits on their length and base diameter are approximately 24 and 8 nm, respectively.[9] These observations are still poorly understood theoretically and determining the optimum conformation of nanocones is crucial if we are to properly understand their growth mechanisms. In this letter we present a model which predicts the dimensions observed for carbon nanocones and explains why the cone angle of 19.2° is preferred.

Physically, the optimum conformation of carbon nanocones is determined by two factors: (1) the competition between energy and entropy and (2) the buckling of the nanocone wall due to pressure induced by surface tension. For a system of carbon nanocones, the free energy is

$$F = E - TS = E_{el} + E_d - T(S_s + S_m), \qquad (1)$$

in which $E_{el}$ is the elastic energy, $S_s$ is the structure entropy, $S_m$ is the mixing entropy,[18–20] and $E_d$ is total defect energy which is composed of the core energy of defects and the interaction energy between defects.[21]

In the continuum limit,[12,18,22–26] the elastic energy of the nanocone (see Fig. 1) can be derived as

$$E_{el} = \pi k_c [\cos^2(\alpha/2)/\sin(\alpha/2)] \ln\{[r + h\tan(\alpha/2)]/r\} + 4\pi(2k_c + k_g), \qquad (2)$$

in which $k_c$ and $k_g$ are the elastic constant and the saddle-splay modulus, respectively, and $r$, $h$, and $\alpha$ are structure parameters (see Fig. 1). On the other hand, using an electrostatic analogy, the energy of defects on a sphere is[15,21]

$$E_d = \pi^2 K_A/18 \sum_{i \neq j}^{n} q_i q_j V(\mathbf{x}_i, \mathbf{x}_j) + \left(\sum_{i=1}^{n} q_i^2\right) E_c, \qquad (3)$$

with the interaction potential between two defects located at positions $\mathbf{x}_i$ and $\mathbf{x}_j$. The potential energy is expressed by

$$V(\mathbf{x}_i, \mathbf{x}_j) = -\ln[(1 - \cos\beta_{ij})/2]/4\pi, \qquad (4)$$

in which $K_A$ is the stiffness constant, $q_i$ is the disclination charge, $\beta_{ij}$ is the geodesic distance, and $E_c$ is the core energy of a disclination.[15,21] For simplicity, we take the average value of $\beta_{ij}$ to calculate the average interaction energy.

By Shannon's formula $S_s = k_B \ln \Omega_s$ the structure entropy is determined by $\Omega_s$, the number of structure states of carbon nanocones for a fixed number of carbon atoms $N$. With the introduction of the chiral indices $(n,m)$, we have[18]

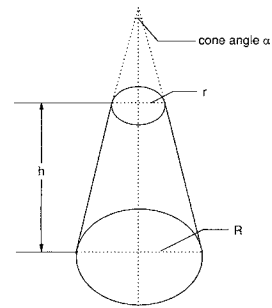

FIG. 1. Demonstration of a carbon nanocone's cross section along the central axis. Two spherical caps are both tangential to the frustum. It can be considered to be frustum with two tangential spherical caps. There are 12 carbon pentagons distributed in the two spherical caps of the cone to form a closed structure. Any pentagon defects exist only at the two ends of a carbon nanocone.

[a]Electronic mail: zhangsl@mail.xjtu.edu.cn

$$\Omega_s(N) = \sum_{i=1}^{6} \Omega_1(\alpha_i, N), \quad (5)$$

with

$$\Omega_1(\alpha_i, N) = \sqrt{3}\pi N \cos^2(\alpha_i/2)/4 + \sqrt{\sqrt{3}\pi N} \cos(\alpha_i/2)/4 - 3. \quad (6)$$

Furthermore, for a nanocone with fixed shape, there is freedom for the 12 carbon pentagons to mix with hexagons on the two ends. This is the origin of the mixing entropy.[18–20] The total number of configurations $\Omega_m$ is given by

$$\Omega_m = \binom{P}{M_1} \times \binom{12-P}{M_2}, \quad (7)$$

where $M_1$ and $M_2$ represent the total number of carbon polygons on the smaller end and the other end, respectively. So, the mixing entropy of a carbon nanocone is

$$S_m = k_B \ln \Omega_m = k_B \ln \binom{P}{2\pi r^2 \sec(\alpha/2)[\sec(\alpha/2) - \tan(\alpha/2)]/2\sigma} + k_B \ln \binom{12-P}{2\pi[r+h\tan(\alpha/2)]^2 \sec(\alpha/2)[\sec(\alpha/2) + \tan(\alpha/2)]/2\sigma}, \quad (8)$$

where $\sigma$ is the area per atom.

Substituting Eqs. (2), (3), (5), and (8) into Eq. (1), the free energy of the carbon nanocone can subsequently be expressed analytically.

It is generally accepted that a carbon nanocone grows from a seed (a curved carbon surface, see the small cap in Fig. 1), which determines the cone angle of the nanocone.[8,9,13] In order to determine the optimum cone angle of carbon nanocones, we should compare the free energy of the five types of seeds for carbon nanocones with the same radius. Since a seed has an open edge, we have to consider the dangling bond energy of carbon atoms at the edge.[9,15,27] By only considering the portion of the equation for the free energy representing the small cap of a carbon nanocone, and adding terms for the dangling bond energy, we have the expression of free energy for the seed as follows:

$$F_r = 2\pi(2k_c + k_g)[1 + \sin(\alpha/2)] + 9.44\pi r/(\sqrt{3}d) + \pi^2 K_A/18 \sum_{i \neq j}^{n} q_i q_j V(\mathbf{x}_i, \mathbf{x}_j) + PE_c$$
$$- k_B T \ln\left(\frac{P}{2\pi r^2 \sec(\alpha/2)[\sec(\alpha/2) - \tan(\alpha/2)]/2\sigma}\right) - k_B T \ln\left\{\sum_{i=1}^{6} [\sqrt{3}\pi N \cos^2(\alpha_i/2)/4 + \sqrt{\sqrt{3}\pi N}\cos(\alpha_i/2)/4 - 3]\right\}, \quad (9)$$

where the second term represents the dangling bond energy (in eV) (Ref. 27) and the synthesis temperature $T=4000$ K.[9,28] This expression for the free energy of a carbon nanocone seed can now be used to understand the observed optimum cone angle of 19.2°.

The free energy $F_r$ of a seed versus the cone angle for the same radius is plotted in Fig. 2. We see that the bigger the cone angle $\alpha$ is, the larger the free energy $F_r$ is. This indicates that the synthesis of a nanocone with big cone angle is more difficult in experiments than that with small cone angle, explaining the preferred cone angle $\alpha=19.2°$ observed experimentally.[9] We also notice that the energy difference between two cone angles is big compared with that of the thermal fluctuation, which is a fraction of an eV. So the thermal fluctuations would have little influence on the optimal conformation of carbon nanocones.

After the seed is formed, the cone angle is fixed at 19.2°. The nanocone begins to grow from the seed towards the larger end. The buckling of the larger end of the carbon nanocone,[29,30] which prevents it from continuously growing to infinity, occurs when

$$2\gamma/R > 3D/R^3. \quad (10)$$

The left term is the pressure induced by the surface tension of the graphitic network applied to the open edge of the nanocone,[31] where $\gamma$ is the surface tension. The right term is the critical pressure needed to avoid buckling, where $D$ is the flexural rigidity of the carbon wall.[30,32] The value of $\gamma$ is temperature dependent and can be estimated as approximately 9.7 dyn/cm.[33] This yields the value of the cutoff radius $R_c=4.5$ nm. It also approximates the cutoff value of the base radius of nanocones observed in experiment (4 nm).[9] This deviation can be understood as the collision of carbon atoms on the carbon wall during growth, which stimulates the collapse of the open end.[8] Correspondingly, the critical height of a carbon nanocone is obtained from the geometrical relation

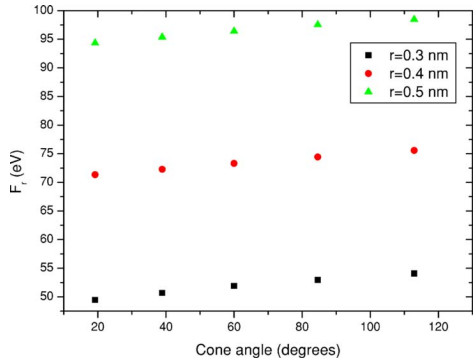

FIG. 2. (Color online) Typical data that describe the dependence of the free energy $F_r$ for a carbon nanocone seed are plotted. It shows that the bigger the cone angle $\alpha$ is, the larger the free energy $F_r$ is.

$$h_c = (R_c - r)/\tan(\alpha/2), \quad (11)$$

from which we see that the height of a carbon nanocone which grows from a seed with radius $r$ cannot exceed $h_c$. We can determine the value of $r$ from the relationship between the free energy of a seed and its radius. Figure 3 shows that the most stable seeds are those with the smallest radius. On the other hand, the geometry of a carbon nanocone and the isolated pentagon rule require that the radii of seeds have a lower bound and the limiting value has the dimension of that of $C_{60}$.[9,17] Therefore we are sure that $r \approx 0.35$ nm, the radius of $C_{60}$. This is confirmed by the experimentally observed sharp tips of nanocones.[9] By Eq. (11) the critical height of a carbon nanocone is $h_c = 24$ nm. Adding the radius of the end, the actual length of the carbon nanocone should be a little larger than 24 nm, which is the experimental value.[9] The consistency of our calculations, that are based on the continuum model, with experimental results indicates the effectiveness of this model in considering the energy of carbon nanocones.

In summary, on the basis of the continuum model, an analytical expression for the free energy of nanocones was proposed and their optimal conformation was studied. The conformation of a carbon nanocone is attributed to both the competition between energy and entropy and the buckling of the nanocone during growth. We found that (1) of the five types of carbon nanocones, the carbon nanocone with cone angle of 19.2° has the minimum free energy; (2) the cutoff radius of nanocones observed in experiment is attributed to the shape transition of buckling; and (3) stable conformation corresponds to a carbon nanocone with $\alpha = 19.2°$, $r = 0.35$ nm, and $h = 24$ nm, which agrees well with experimental observation. Our study of the optimal conformation of carbon nanocones is conducive to understanding the growing mechanism of carbon nanocone, which is beneficial for experimental observation and controllable synthesis.

The authors thank Jian Li, Lei Zhang, Run Liang, Yachao Liu, and Xianjun Zuo for helpful discussions. This work was supported by the NSF of China Grant Nos. 10374075 and 10547135.

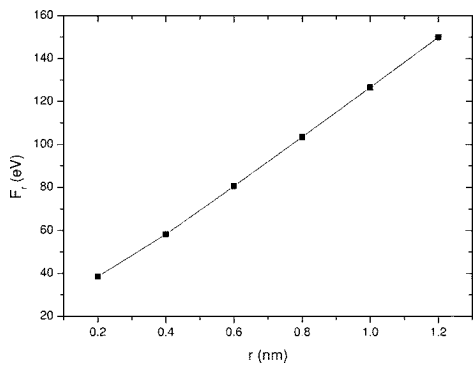

FIG. 3. Free energy of seeds varies with radii. The cone angle of seeds is 19.2°, i.e., there are five pentagons on the seeds.